\documentclass[aps,prb,reprint,superscriptaddress]{revtex4-2}

\usepackage{amssymb}
\usepackage{mathtools}
\usepackage{hyperref}
\usepackage{bm}
\usepackage{tikz}
\usetikzlibrary{decorations.pathmorphing}

\newcommand{\ket}[1]{\vert #1 \rangle}

\begin{document}

\title{Non-invertible duality and symmetry topological order of one-dimensional lattice models with spatially modulated symmetry}

\author{Donghae Seo}
\affiliation{Department of Physics, Pohang University of Science and Technology, Pohang 37673, Republic of Korea}
\affiliation{Center for Artificial Low Dimensional Electronic Systems, Institute for Basic Science, Pohang 37673, Republic of Korea}

\author{Gil Young Cho}
\email{gilyoungcho@kaist.ac.kr}
\affiliation{Department of Physics, Korea Advanced Institute of Science and Technology, Daejeon 34141, Republic of Korea}
\affiliation{Center for Artificial Low Dimensional Electronic Systems, Institute for Basic Science, Pohang 37673, Republic of Korea}
\affiliation{Asia-Pacific Center for Theoretical Physics, Pohang 37673, Republic of Korea}

\author{Robert-Jan Slager}
\affiliation{Department of Physics and Astronomy, University of Manchester, Oxford Road, Manchester M13 9PL, United Kingdom}
\affiliation{TCM Group, Cavendish Laboratory, University of Cambridge, JJ Thomson Avenue, Cambridge CB3 0HE, United Kingdom}

\date{\today}

\begin{abstract}
    We investigate the interplay between self-duality and spatially modulated symmetry of generalized $N$-state clock models, which include the transverse-field Ising model and ordinary $N$-state clock models as special cases. The spatially modulated symmetry of the model becomes trivial when the model's parameters satisfy a specific number-theoretic relation. We find that the duality is non-invertible when the spatially modulated symmetry remains nontrivial, and show that this non-invertibility is resolved by introducing a generalized $\mathbb{Z}_N$ toric code, which manifests ultraviolet/infrared mixing, as the bulk topological order. In this framework, the boundary duality transformation corresponds to the boundary action of a bulk symmetry transformation, with the endpoint of the bulk symmetry defect realizing the boundary duality defect. Our results illuminate not only a holographic perspective on dualities but also a relationship between spatially modulated symmetry and ultraviolet/infrared mixing in one higher dimension.
\end{abstract}

\maketitle

\section{Introduction}

Dualities are powerful tools for analyzing non-perturbative problems. Broadly speaking, a duality represents a nontrivial correspondence between seemingly distinct interpretations. Upon relating both sides to a single theory, dualities may map a challenging problem into a simpler one---for instance, transforming a strongly coupled system into a weakly coupled one. Since they are non-perturbative, dualities are invaluable theoretical tools for studying physical systems that are difficult to approach directly and have proven effective in areas ranging from strongly interacting matter and conformal field theories to spin systems and lattice gauge theories~\cite{kramers1941statistics,kogut1979introduction,witten1998anti,maldacena1999large,seiberg2016duality,beekman2017dual,ruegg2024dualities}.

Duality mappings are evidently governed by symmetry constraints. While symmetry is a cornerstone of physics, the concept has been generalized in recent years. In particular, one can define higher-form symmetries that act on higher-codimensional objects~\cite{gaiotto2015generalized,gomes2023introduction,bhardwaj2024lectures,luo2024lecture}, such as Wilson lines. Although non-invertible symmetries~\cite{shao2024what,schafernameki2024ictp} have only recently come into focus, they appear surprisingly frequently in the context of duality. A notable example is the well-studied Kramers-Wannier duality~\cite{kramers1941statistics}, where the systems which are self-dual under it possess a non-invertible symmetry~\cite{shao2024what,schafernameki2024ictp}. These non-invertible transformations occur because certain duality mappings reconfigure the model's topological sectors. For instance, in the one-dimensional lattice model with $\mathbb{Z}_2$ symmetry, the Kramers-Wannier duality transformation maps the $\mathbb{Z}_2$-odd sub-Hilbert space to the defect sector. Since this defect sector does not belong to the original untwisted Hilbert space, the Kramers-Wannier duality transformation has a nontrivial kernel in the original space. Thus, by incorporating all topological sectors, these non-invertible dualities can be promoted to isometries~\cite{lootens2023dualities,lootens2023low,lootens2024dualities}.

The perspectives discussed above benefit further from the notion of the symmetry/topological-order correspondence, also known as the topological holographic principle~\cite{moradi2023topological,chatterjee2023symmetry}. In this framework, each symmetry of a $d$-dimensional system uniquely determines a $(d+1)$-dimensional topological order, often called the symmetry topological order. This holographic perspective has been demonstrated through several concrete examples~\cite{chatterjee2023holographic,inamura2023symmetry,vanhove2024duality}, yet considerable potential for further investigation remains. In particular, a promising direction for exploring the symmetry/topological-order correspondence is to apply this formalism to generalized symmetries. 

Interest in spatially modulated symmetries, whose symmetry generators vary across space, has been recently growing~\cite{sala2022dynamics,delfino2023fractons,han2024topological,lam2024classification,sala2024exotic,pace2024gauging}. Given this interest, unifying spatially modulated symmetries within the framework of topological holography would be valuable; however, few studies currently address this topic. For instance, a holographic perspective on certain dipole symmetries, which is a specific category of spatially modulated symmetry, has been explored, leading to symmetry topological orders enriched by spacetime symmetries~\cite{pace2024spt}. Nonetheless, a comprehensive holographic framework for general spatially modulated symmetries remains an open area of research.

In this work, we adopt the topological holography perspective to analyze one-dimensional lattice models with a certain class of parameter-dependent spatially modulated symmetry. Specifically, we examine the exponential symmetries of generalized $N$-state clock models~\cite{hu2023spontaneous}. Notably, our results also extend to the transverse-field Ising model and standard $N$-state clock models, which emerge as special cases when the model parameters are appropriately chosen. These generalized $N$-state clock models exhibit self-duality, and we find that this duality can be either invertible or non-invertible. In particular, this duality is non-invertible when the spatially modulated symmetry is nontrivial; however, the symmetry can become trivial when certain number-theoretic conditions on the model's parameters are satisfied. Moreover, this non-invertibility can be resolved by constraining the Hilbert space of the model, though this introduces a non-invertible gravitational anomaly~\cite{ji2022unified}, necessitating a bulk topological order. We demonstrate that the generalized $\mathbb{Z}_N$ toric code \cite{watanabe2023ground} can serve as this bulk topological order. Within this framework, the boundary duality transformation corresponds to the action of a bulk symmetry transformation at the boundary, with the endpoint of the bulk symmetry defect realizing the boundary duality defect.

Our results suggest several general avenues for investigation. First, we identify an intriguing connection between duality and symmetry: the presence of a spatially modulated symmetry implies a nontrivial kernel of the duality transformation map. Second, we illustrate the bulk-boundary relationship between spatially modulated symmetry and ultraviolet/infrared (UV/IR) mixing---the phenomenon where a system's infrared properties are influenced by its ultraviolet properties~\cite{minwalla2000noncommutative}. A prominent example of UV/IR mixing in lattice models is the generalized $\mathbb{Z}_N$ toric code model, whose topological order depends on the size of the system. Recent studies however show growing interest in more generic UV/IR mixing within lattice models~\cite{gorantla2021low,rudelius2021fractons,you2022fractonic,kim2024unveiling}. Finally, we provide an explicit and unconventional example of the holographic framework for dualities. Specifically, dualities can be viewed as symmetry transformations in a topological order within one higher dimension. 

The main content of this paper is organized as follows. In Sec.~\ref{sec:non-invertible_duality}, we introduce the generalized $N$-state clock model and formulate its duality in terms of a linear map. By analyzing the kernel of this map, we reveal that the duality's non-invertibility is linked to the model's spatially modulated symmetry. In Sec.~\ref{sec:boundary_constraint}, we discuss a constraint imposed on the boundary of the generalized $\mathbb{Z}_N$ toric code, which coincides with the constraint needed on the Hilbert space of the generalized $N$-state clock model to render its duality invertible. In Sec.~\ref{sec:bulk-boundary_correspondence}, we demonstrate that the boundary duality transformation can be implemented via the electric-magnetic self-duality transformation of the bulk, with the endpoint of the bulk duality defect at the boundary identified as the boundary duality defect. Finally, we conclude in Sec.~\ref{sec:conclusions}.

\section{Non-invertible duality of generalized \texorpdfstring{$N$}{N}-state clock model}
\label{sec:non-invertible_duality}

We first introduce the generalized $N$-state clock model constructed by Hu and Watanabe \cite{hu2023spontaneous} and discuss a non-invertible duality of the model. This non-invertible duality is a generalization of the familiar Kramer-Wannier duality in the transverse-field Ising model \cite{kramers1941statistics}. We then establish a connection between the duality and a spatially modulated symmetry of the model. 

The generalized $N$-state clock model is defined on a one-dimensional lattice with $L$ sites, where each lattice site is associated with a local Hilbert space isomorphic to $\mathbb{C}^N$. The Hamiltonian is given by
\begin{equation}
    H = -\frac{1}{2} \sum_{i=1}^L \left[(Z_i^{-a} Z_{i+1} + \mathrm{h.c.}) + g(X_i + \mathrm{h.c.})\right],
\label{1d_model}
\end{equation}
where $a$ is an integer parameter such that $1 \leq a < N$. Here, $X_i$ and $Z_i$ are the generalized $N$-level Pauli operators acting on the $i$-th site. These operators satisfy $Z_i X_i = \omega X_i Z_i$, where $\omega = e^{2\pi i/N}$, and commute with each other at different sites. A periodic boundary condition is imposed, where $Z_{L+1} \equiv Z_1$.

The transverse-field Ising model and ordinary $N$-state clock models are special cases of Eq.~\eqref{1d_model}. Specifically, the transverse-field Ising model is recovered by setting $N = 2$ and $a = 1$, while the ordinary clock models correspond to the cases where $a = 1$ and $N > 2$ is arbitrary.

The duality transformation that we will focus on in this work is defined as
\begin{equation}
    Z_i^{a} Z_{i+1}^{-1} \mapsto \tilde{X}_{i+1}, \quad X_i \mapsto \tilde{Z}_i^{-1} \tilde{Z}_{i+1}^a \quad \forall i,
\end{equation}
under which Eq.~\eqref{1d_model} transforms into
\begin{equation}
    \tilde{H} = -\frac{1}{2} \sum_{i=1}^L \left[(\tilde{X}_i + \mathrm{h.c.}) + g(\tilde{Z}_i^{-1} \tilde{Z}_{i+1}^a + \mathrm{h.c.})\right].
\end{equation}
When $g = 1$, the dual Hamiltonian maps back to the original Hamiltonian Eq.~\eqref{1d_model} up to the spatial inversion, i.e, $i \to L-i$. The spatial inversion is redundant and has been absent in the conventional cases of $a=1$ or in the Ising model. However, it is necessary if $a \neq 1$.

We focus on the case of $\gcd(a^L-1, N) \neq 1$, where the model exhibits a spatially modulated $\mathbb{Z}_{\gcd(a^L-1, N)}$ symmetry and a non-invertible duality. {This symmetry corresponds an exponential symmetry and is the discrete spin-rotation symmetry generated by}
\begin{equation}
    \eta^n = \left(\prod_{i=1}^L X_i^{a^{i-1}}\right)^n, 
\label{modulatedSym}
\end{equation}
where $n = N / \gcd(a^L-1, N)$. Under the duality transformation, Eq.~\eqref{modulatedSym} is mapped to 
\begin{equation}
    \tilde{\eta}^n \equiv \left(\prod_{i=1}^L (\tilde{Z}_i^{-1} \tilde{Z}_{i+1}^a)^{a^{i-1}}\right)^n = \tilde{Z}_1^{n(a^L-1)}.
\end{equation}
Interestingly, $\tilde{\eta}^n$ becomes the identity operator when $\gcd(a^L-1, N) \neq 1$:
\begin{equation}
    \tilde{\eta}^n = \tilde{Z}_1^{n(a^L-1)} = \tilde{Z}_1^{mN} = 1,
\end{equation}
where $m$ is an integer such that $m = N / n(a^L - 1)$. If this duality map could be implemented by an invertible operator, $\eta^n$ would also have been the identity operator, which is not the case here. Hence, when the generalized $N$-state clock model exhibits this spatially modulated symmetry, the duality map cannot be implemented by an invertible operator, indicating that it is non-invertible.

The non-invertible nature of the duality can be manifested from the nontrivial kernel of the duality map. To analyze the kernel, we introduce a linear map that implements the duality transformation:
\begin{equation}
    \mathsf{D} \ket{\mathbf{s}} = \frac{1}{N^{L/2}} \sum_{\tilde{\mathbf{s}}} \omega^{\sum_{i=1}^L (a s_i - s_{i+1}) \tilde{s}_{i+1}} \ket{\tilde{\mathbf{s}}},
    \label{Duality}
\end{equation}
where $\mathbf{s}$ and $\tilde{\mathbf{s}}$ are $L$-dimensional $\mathbb{Z}_N$ vectors, and the summation is over all such vectors. The map $\mathsf{D}$ is a map from the original Hilbert space to the dual Hilbert space, i.e., $\ket{\mathbf{s}}$ in the left-hand side of Eq.~\eqref{Duality} lives in the original Hilbert space but $\ket{\tilde{\mathbf{s}}}$ in the right-hand side of Eq.~\eqref{Duality} is defined in the dual Hilbert space. Since $\mathsf{D}$ is a linear map, Eq.~\eqref{Duality} determines the duality transformation for any state. We note that a similar formulation has been proposed for describing the Kramers-Wannier duality and some of its generalizations in Ref.~\cite{yan2024generalized}. 

The fact that $\mathsf{D}$ implements the anticipated duality transformation can be demonstrated as follows:
\begin{equation}
    \begin{aligned}
        \mathsf{D} Z_I^a Z_{I+1}^{-1} \ket{\mathbf{s}} &= \frac{\omega^{as_I - s_{I+1}}}{N^{L/2}} \sum_{\tilde{\mathbf{s}}} \omega^{\sum_{i=1}^L (as_i - s_{i+1}) \tilde{s}_{i+1}} \ket{\tilde{\mathbf{s}}} \\
        &= \frac{1}{N^{L/2}} \sum_{\tilde{\mathbf{s}}} \omega^{\sum_{i=1}^L (as_i - s_{i+1}) \tilde{s}_{i+1} + as_I - s_{I+1}} \ket{\tilde{\mathbf{s}}} \\
        &= \tilde{X}_{I+1} \mathsf{D} \ket{\mathbf{s}}, \\
        \mathsf{D} X_I \ket{\mathbf{s}} &= \frac{1}{N^{L/2}} \sum_{\tilde{\mathbf{s}}} \omega^{\sum_{i=1}^L (as_i - s_{i+1}) \tilde{s}_{i+1} - \tilde{s}_I + a \tilde{s}_{I+1}} \ket{\tilde{\mathbf{s}}} \\
        &= \tilde{Z}_I^{-1} \tilde{Z}_{I+1}^a \mathsf{D} \ket{\mathbf{s}}.
    \end{aligned}
\end{equation}
We thus observe that $\mathsf{D}$ indeed implements the duality transformation.

The kernel of $\mathsf{D}$ becomes more apparent when we rewrite it as 
\begin{equation}
    \mathsf{D} \ket{\mathbf{s}} = \frac{1}{N^{L/2}} \sum_{\tilde{\mathbf{s}}} \omega^{\mathbf{s}^T \delta \tilde{\mathbf{s}}} \ket{\tilde{\mathbf{s}}},
\end{equation}
where 
\begin{equation}
    \delta = 
    \begin{pmatrix}
        0 & 0 & 0 & 0 & \cdots & a \\
        a & 0 & 0 & 0 & \cdots & -1 \\
        -1 & a & 0 & 0 & \cdots & 0 \\
        0 & -1 & a & 0 & \cdots & 0 \\
        \vdots & \vdots & \vdots & \vdots & \ddots & \vdots \\
        0 & 0 & 0 & 0 & \cdots & 0
    \end{pmatrix}
\end{equation}
is an $(L \times L)$ integer matrix whose elements are defined modulo $N$. If $\delta$ has a nontrivial cokernel, then the associated duality is non-invertible. To see this, let an $L$-dimensional integer covector $\mathbf{v}^T$ be in the cokernel of $\delta$. In this case, we have $\mathsf{D} \ket{\mathbf{s}} = \mathsf{D} \ket{\mathbf{s} + \mathbf{v}}$, which implies that $\mathsf{D} (\ket{\mathbf{s}} - \ket{\mathbf{s} + \mathbf{v}}) = 0$. Therefore, $\mathsf{D}$ has a nontrivial kernel, indicating that the duality transformation is non-invertible.

The cokernel of $\delta$ encodes the spatially modulated symmetry of the generalized $N$-state clock model. To see this, we note that the addition of a covector $\mathbf{v}$ to $\mathbf{s}$ corresponds to the rotation of each $i$-th spin by $v_i$, the $i$-th element of $\mathbf{v}$. Utilizing this, we can convert the action of the spatially-modulated symmetry Eq.~\eqref{modulatedSym} to the addition of the covector $n\bm{\eta}^T = n(1, a, a^2, \cdots, a^{L-2}, a^{L-1})$. Notably, this covector is indeed a cokernel of $\delta$ because
\begin{equation}
    \begin{aligned}
        n \bm{\eta}^T \delta &= \left(0, 0, 0, \cdots, n(1-a^L), 0\right) \\
        &\equiv (0, 0, 0, \cdots, 0, 0) \mod N.
    \end{aligned}
\end{equation}
Thus, the spatially modulated symmetry of the generalized $N$-state clock model is in a one-to-one correspondence with the cokernel of $\delta$, and consequently in the kernel of $\mathsf{D}$.

\section{Boundary constraint of generalized \texorpdfstring{$\mathbb{Z}_N$}{Z\_N} toric code}
\label{sec:boundary_constraint}

As a next step, we now demonstrate that the non-invertible duality in the generalized $N$-state clock model can be upgraded to an invertible symmetry within a constrained Hilbert space of the same model. Due to the constraint on the Hilbert space, the model cannot be formulated on a one-dimensional lattice; instead, it is realized as a boundary theory of the generalized $\mathbb{Z}_N$ toric code.

To eliminate the kernel of $\mathsf{D}$, we restrict the Hilbert space to the symmetric subspace under $\eta^n$. The restriction is implemented by the projector 
\begin{equation}
    \mathcal{P}(N,L,a) \coloneqq \frac{1}{\sqrt{\gcd(a^L-1,N)}} \sum_{k=0}^{\gcd(a^L-1,N)-1} \eta^{nk}.
\end{equation}
We note that any state within the symmetric subspace is invariant under the symmetry operator as
\begin{equation}
    \begin{aligned}
        \eta^n \mathcal{P}(N,L,a) &= \frac{1}{\sqrt{\gcd(a^L-1,N)}} \sum_{k=0}^{\gcd(a^L-1,N)-1} \eta^{n(k + 1)} \\
        &= \frac{\sum_{k=1}^{\gcd(a^L-1,N)-1} \eta^{nk} + \eta^N}{\sqrt{\gcd(a^L-1,N)}} \\
        &= \mathcal{P}(N,L,a).
    \end{aligned}
\end{equation}
In this symmetric subspace, $\mathsf{D}$ no longer has a nontrivial kernel and thus indeed becomes invertible.

It is important to note that the constrained system cannot be implemented as a lattice model in the same dimension. Instead, it can be realized as a boundary theory of a topologically ordered lattice model \cite{ji2022unified}. Specifically, we will demonstrate that the boundary theory of the generalized $\mathbb{Z}_N$ toric code \cite{watanabe2023ground} realizes the generalized $N$-state clock model with its Hilbert space restricted to the symmetric subspace.

We begin by recalling that the generalized $\mathbb{Z}_N$ toric code \cite{watanabe2023ground} is defined on a square lattice, where each $N$-level qudit is located at each edge. The Hamiltonian of the model is given by 
\begin{equation}
    H = - \frac{1}{2} \sum_{v \in \mathcal{V}} (A_v + \mathrm{h.c.}) - \frac{1}{2} \sum_{p \in \mathcal{P}} (B_p + \mathrm{h.c.}),
\end{equation}
where 
\begin{equation}
    A_v \equiv 
    \begin{tikzpicture}[baseline={([yshift=-.8ex]current bounding box.center)}]
        \draw (0,0) -- (1,0); \draw (0.5,-0.5) -- (0.5,0.5); \node[anchor=east] at (0,0) {$X$}; \node[anchor=north] at (0.5,-0.5) {$X$}; \node[anchor=west] at (1,0) {$X^{-a_1}$}; \node[anchor=south] at (0.5,0.5) {$X^{-a_2}$}; \node[anchor=south west] at (0.5,0) {$v$};
    \end{tikzpicture}, \quad 
    B_p \equiv
    \begin{tikzpicture}[baseline={([yshift=-1ex]current bounding box.center)}]
        \draw (0,0) -- (1,0); \draw (1,0) -- (1,1); \draw (1,1) -- (0,1); \draw (0,1) -- (0,0); \node[anchor=east] at (0,0.5) {$Z^{-a_1}$}; \node[anchor=north] at (0.5,0) {$Z^{a_2}$}; \node[anchor=west] at (1,0.5) {$Z$}; \node[anchor=south] at (0.5,1) {$Z^{-1}$}; \node at (0.5,0.5) {$p$};
    \end{tikzpicture}.
\end{equation}
Here, $a_1$ and $a_2$ are the two integer parameters such that $1 \leq a_1 < N$ and $1 \leq a_2 < N$. We set $a_1 = a_2 \equiv a$ to make the boundary theory of the generalized $\mathbb{Z}_N$ toric code match with the generalized $N$-state clock model given by Eq.~\eqref{1d_model}.

\begin{figure}
    \centering
    \includegraphics[]{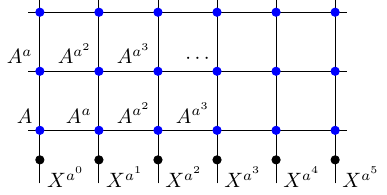}
    \caption{The rough boundary of generalized $\mathbb{Z}_N$ toric code and boundary constraint from a bulk conservation law. Blue dots represent operators acting on adjacent qudits, while black dots denote the boundary qudits constrained by the bulk conservation law.}
    \label{fig:rough_boundary}
\end{figure}

Consider the bottom rough boundary of the generalized $\mathbb{Z}_N$ toric code defined on a semi-infinite cylinder, as illustrated in Fig.~\ref{fig:rough_boundary}. A bulk conservation law is defined by multiplying certain powers of $A_v$ at each vertex such that the operators acting on the bulk qudits cancel each other out. Figure~\ref{fig:rough_boundary} depicts such a bulk conservation law. For the bulk operators to cancel along the periodic boundary of the cylinder, $N$ must divide $(a^L - 1)$, which implies $n = N / \gcd(a^L - 1, N) = 1$.

In cases where this condition is not met, we can modify the bulk conservation law by taking the $n$-th power of each stabilizer, where $n$ is given by $n = N / \gcd(a^L-1, N)$. Then, the bulk operators cancel entirely because $X^{-n(a^L - 1)} = X^N = 1$. This bulk conservation law imposes a nonlocal constraint at the boundary 
\begin{equation}
    \left(\prod_{i=1}^L X_i^{a^{i-1}}\right)^n = 1
\end{equation}
where $L$ is the number of the qudits at the boundary and $i$ labels the boundary qudits. Notably, this is the same constraint imposed on the Hilbert space of the the generalized $N$-state clock model to render its duality invertible. 

The constrained generalized $N$-state clock model can be realized as a boundary theory of the generalized $\mathbb{Z}_N$ toric code. The boundary operators can be chosen as those commute with the bulk stabilizers. A choice of the boundary operators is given by 
\begin{equation}
    \bar{A}_i \equiv
    \begin{tikzpicture}[baseline={([yshift=-.8ex]current bounding box.center)}]
        \draw (0,0) -- (0,1); \filldraw[black] (0,0.5) circle (2pt); \node[anchor=east] at (0,0.5) {$X$}; \node[anchor=west] at (0,0.5) {$i$};
    \end{tikzpicture}, \quad 
    \bar{B}_{i,i+1} \equiv 
    \begin{tikzpicture}[baseline={([yshift=-2.5ex]current bounding box.center)}]
        \draw (0,0) -- (0,1); \draw (0,1) -- (1,1); \draw (1,1) -- (1,0);
        \filldraw[black] (0,0.5) circle (2pt); \filldraw[black] (0.5,1) circle (2pt); \filldraw[black] (1,0.5) circle (2pt); 
        \node[anchor=west] at (0,0.5) {$i$}; \node[anchor=west] at (1,0.5) {$i+1$};
        \node[anchor=east] at (0,0.5) {$Z^{-a}$}; \node[anchor=east] at (1,0.5) {$Z$}; \node[anchor=south] at (0.5,1) {$Z^{-1}$};
    \end{tikzpicture},
\end{equation}
where the boundary qudits are labeled by $i$. These boundary operators generate the same algebra as that generated by the terms, such as $X_i$ and $Z_i^{-a} Z_{i+1}$, in the Hamiltonian Eq.~\eqref{1d_model} of the generalized $N$-state clock model. Therefore, the Hamiltonian 
\begin{equation}
    H = - \frac{1}{2} \sum_{i=1}^L \left[(\bar{B}_{i,i+1} + \mathrm{h.c.}) + g(\bar{A}_i + \mathrm{h.c.})\right]
\end{equation}
has the same spectrum, within the boundary Hilbert space, as those of the constrained generalized $N$-state clock model.

On the other hand, consider the top smooth boundary of the generalized $\mathbb{Z}_N$ toric code. A choice of the boundary operators is given by 
\begin{equation}
    \tilde{A}_i \equiv 
    \begin{tikzpicture}[baseline={([yshift=-.8ex]current bounding box.center)}]
        \draw (0,0) -- (1,0); \draw (0.5,0) -- (0.5,-0.5);
        \filldraw[black] (0.5,0) circle (2pt); \node[anchor=south] at (0.5,0) {$\tilde{i}$};
        \node[anchor=east] at (0,0) {$X$}; \node[anchor=west] at (1,0) {$X^{-a}$}; \node[anchor=north] at (0.5,-0.5) {$X$};
    \end{tikzpicture}, \quad 
    \tilde{B}_{i,i+1} \equiv 
    \begin{tikzpicture}[baseline={([yshift=0ex]current bounding box.center)}]
        \draw (0,0) -- (1,0); \filldraw[black] (0,0) circle (2pt); \filldraw[black] (1,0) circle (2pt);
        \node[anchor=north] at (0,0) {$\tilde{i}$}; \node[anchor=north] at (1,0) {$\tilde{i}+1$};
        \node[anchor=south] at (0.5,0) {$Z^{-1}$};
    \end{tikzpicture},
\end{equation}
where $\tilde{i}$ labels the degrees of freedom at the boundary. Here, the boundary degrees of freedom are intentionally chosen so that they represent the dual sites. The boundary operators satisfy the following algebra
\begin{equation}
    \tilde{A}_{\tilde{i}} \tilde{B}_{\tilde{i},\tilde{i}+1} = \omega^{-a} \tilde{B}_{\tilde{i},\tilde{i}+1} \tilde{A}_{\tilde{i}},~ \tilde{A}_{\tilde{i}+1} \tilde{B}_{\tilde{i},\tilde{i}+1} = \omega \tilde{B}_{\tilde{i},\tilde{i}+1} \tilde{A}_{\tilde{i}}
\end{equation}
and any other pair of them commute. Importantly, these boundary operators generated the same operator algebra as that generated by the terms, such as $X_{\tilde{i}}$ and $Z_{\tilde{i}}^{-1} Z_{\tilde{i}+1}$, in the dual model of the generalized $N$-state clock model.

\section{Bulk symmetry/boundary duality correspondence}
\label{sec:bulk-boundary_correspondence}

We now explore the connection between the non-invertible duality in the generalized $N$-state clock model and a symmetry of the generalized $\mathbb{Z}_N$ toric code. The generalized $\mathbb{Z}_N$ toric code possesses the electric-magnetic self-duality, which corresponds to a $\mathbb{Z}_2$ symmetry of the model. The corresponding symmetry transformation is implemented through the following three steps.

As the first step, we apply $U_Y$ and $U_Y^\dagger$, which acts on the Pauli operators such that $U_Y X U_Y^\dagger = Z$ and $U_Y Z U_Y^\dagger = X^\dagger$, to the qudits on the horizontal and vertical edges, respectively. Under this transformation, the bulk stabilizers are mapped as follows:
\begin{equation}
\begin{aligned}
    \begin{tikzpicture}[baseline={([yshift=-.8ex]current bounding box.center)}]
        \draw (0,0) -- (1,0); \draw (0.5,-0.5) -- (0.5,0.5); 
        \node[anchor=east] at (0,0) {$X$}; \node[anchor=north] at (0.5,-0.5) {$X$}; \node[anchor=west] at (1,0) {$X^{-a_1}$}; \node[anchor=south] at (0.5,0.5) {$X^{-a_2}$};
    \end{tikzpicture}
    &\mapsto
    \begin{tikzpicture}[baseline={([yshift=-.8ex]current bounding box.center)}]
        \draw (0,0) -- (1,0); \draw (0.5,-0.5) -- (0.5,0.5); 
        \node[anchor=east] at (0,0) {$Z$}; \node[anchor=north] at (0.5,-0.5) {$Z^{-1}$}; \node[anchor=west] at (1,0) {$Z^{-a_1}$}; \node[anchor=south] at (0.5,0.5) {$Z^{a_2}$};
    \end{tikzpicture}, \\
    \begin{tikzpicture}[baseline={([yshift=-1ex]current bounding box.center)}]
        \draw (0,0) -- (1,0); \draw (1,0) -- (1,1); \draw (1,1) -- (0,1); \draw (0,1) -- (0,0); 
        \node[anchor=east] at (0,0.5) {$Z^{-a_1}$}; \node[anchor=north] at (0.5,0) {$Z^{a_2}$}; \node[anchor=west] at (1,0.5) {$Z$}; \node[anchor=south] at (0.5,1) {$Z^{-1}$};
    \end{tikzpicture}
    &\mapsto
    \begin{tikzpicture}[baseline={([yshift=-1ex]current bounding box.center)}]
        \draw (0,0) -- (1,0); \draw (1,0) -- (1,1); \draw (1,1) -- (0,1); \draw (0,1) -- (0,0); 
        \node[anchor=east] at (0,0.5) {$X^{-a_1}$}; \node[anchor=north] at (0.5,0) {$X^{-a_2}$}; \node[anchor=west] at (1,0.5) {$X$}; \node[anchor=south] at (0.5,1) {$X$};
    \end{tikzpicture}.
\end{aligned}
\end{equation}
Next, we consider the dual of the lattice:
\begin{equation}
\begin{aligned}
    \begin{tikzpicture}[baseline={([yshift=-.8ex]current bounding box.center)}]
        \draw (0,0) -- (1,0); \draw (0.5,-0.5) -- (0.5,0.5); 
        \node[anchor=east] at (0,0) {$Z$}; \node[anchor=north] at (0.5,-0.5) {$Z^{-1}$}; \node[anchor=west] at (1,0) {$Z^{-a_1}$}; \node[anchor=south] at (0.5,0.5) {$Z^{a_2}$};
    \end{tikzpicture}
    &\mapsto
    \begin{tikzpicture}[baseline={([yshift=-1ex]current bounding box.center)}]
        \draw (0,0) -- (1,0); \draw (1,0) -- (1,1); \draw (1,1) -- (0,1); \draw (0,1) -- (0,0); 
        \node[anchor=east] at (0,0.5) {$Z$}; \node[anchor=north] at (0.5,0) {$Z^{-1}$}; \node[anchor=west] at (1,0.5) {$Z^{-a_1}$}; \node[anchor=south] at (0.5,1) {$Z^{a_2}$};
    \end{tikzpicture}, \\
    \begin{tikzpicture}[baseline={([yshift=-1ex]current bounding box.center)}]
        \draw (0,0) -- (1,0); \draw (1,0) -- (1,1); \draw (1,1) -- (0,1); \draw (0,1) -- (0,0); 
        \node[anchor=east] at (0,0.5) {$X^{-a_1}$}; \node[anchor=north] at (0.5,0) {$X^{-a_2}$}; \node[anchor=west] at (1,0.5) {$X$}; \node[anchor=south] at (0.5,1) {$X$};
    \end{tikzpicture}
    &\mapsto 
    \begin{tikzpicture}[baseline={([yshift=-.8ex]current bounding box.center)}]
        \draw (0,0) -- (1,0); \draw (0.5,-0.5) -- (0.5,0.5);
        \node[anchor=east] at (0,0) {$X^{-a_1}$}; \node[anchor=north] at (0.5,-0.5) {$X^{-a_2}$}; \node[anchor=west] at (1,0) {$X$}; \node[anchor=south] at (0.5,0.5) {$X$};
    \end{tikzpicture}.
\end{aligned}
\end{equation}
Finally, we rotate the lattice by $180$ degrees:
\begin{equation}
\begin{aligned}
    \begin{tikzpicture}[baseline={([yshift=-1ex]current bounding box.center)}]
        \draw (0,0) -- (1,0); \draw (1,0) -- (1,1); \draw (1,1) -- (0,1); \draw (0,1) -- (0,0); 
        \node[anchor=east] at (0,0.5) {$Z$}; \node[anchor=north] at (0.5,0) {$Z^{-1}$}; \node[anchor=west] at (1,0.5) {$Z^{-a_1}$}; \node[anchor=south] at (0.5,1) {$Z^{a_2}$};
    \end{tikzpicture}
    &\mapsto 
    \begin{tikzpicture}[baseline={([yshift=-1ex]current bounding box.center)}]
        \draw (0,0) -- (1,0); \draw (1,0) -- (1,1); \draw (1,1) -- (0,1); \draw (0,1) -- (0,0); 
        \node[anchor=east] at (0,0.5) {$Z^{-a_1}$}; \node[anchor=north] at (0.5,0) {$Z^{a_2}$}; \node[anchor=west] at (1,0.5) {$Z$}; \node[anchor=south] at (0.5,1) {$Z^{-1}$};
    \end{tikzpicture}, \\
    \begin{tikzpicture}[baseline={([yshift=-.8ex]current bounding box.center)}]
        \draw (0,0) -- (1,0); \draw (0.5,-0.5) -- (0.5,0.5);
        \node[anchor=east] at (0,0) {$X^{-a_1}$}; \node[anchor=north] at (0.5,-0.5) {$X^{-a_2}$}; \node[anchor=west] at (1,0) {$X$}; \node[anchor=south] at (0.5,0.5) {$X$};
    \end{tikzpicture}
    &\mapsto 
    \begin{tikzpicture}[baseline={([yshift=-.8ex]current bounding box.center)}]
        \draw (0,0) -- (1,0); \draw (0.5,-0.5) -- (0.5,0.5);
        \node[anchor=east] at (0,0) {$X$}; \node[anchor=north] at (0.5,-0.5) {$X$}; \node[anchor=west] at (1,0) {$X^{-a_1}$}; \node[anchor=south] at (0.5,0.5) {$X^{-a_2}$};
    \end{tikzpicture}.
\end{aligned}
\end{equation}
As a result, the vertex and plaquette stabilizers are interchanged, while the overall Hamiltonian remains invariant. This thus finalized the transformation.

Under the bulk symmetry action, the boundary operators associated with the rough and smooth boundaries are exchanged:
\begin{equation}
\begin{aligned}
    \begin{tikzpicture}[baseline={([yshift=-.8ex]current bounding box.center)}]
        \draw (0,0) -- (1,0); \draw (0.5,0) -- (0.5,-0.5);
        \filldraw[black] (0.5,0) circle (2pt); \node[anchor=south] at (0.5,0) {$\tilde{i}$};
        \node[anchor=east] at (0,0) {$X$}; \node[anchor=west] at (1,0) {$X^{-a}$}; \node[anchor=north] at (0.5,-0.5) {$X$};
    \end{tikzpicture} &\mapsto 
    \begin{tikzpicture}[baseline={([yshift=-2.5ex]current bounding box.center)}]
        \draw (0,0) -- (0,1); \draw (0,1) -- (1,1); \draw (1,1) -- (1,0);
        \filldraw[black] (0,0.5) circle (2pt); \filldraw[black] (0.5,1) circle (2pt); \filldraw[black] (1,0.5) circle (2pt); 
        \node[anchor=west] at (0,0.5) {$i$}; \node[anchor=west] at (1,0.5) {$i+1$};
        \node[anchor=east] at (0,0.5) {$Z^{-a}$}; \node[anchor=east] at (1,0.5) {$Z$}; \node[anchor=south] at (0.5,1) {$Z^{-1}$};
    \end{tikzpicture}, \\
    \begin{tikzpicture}[baseline={([yshift=0ex]current bounding box.center)}]
        \draw (0,0) -- (1,0); \filldraw[black] (0,0) circle (2pt); \filldraw[black] (1,0) circle (2pt);
        \node[anchor=north] at (0,0) {$\tilde{i}$}; \node[anchor=north] at (1,0) {$\tilde{i}+1$};
        \node[anchor=south] at (0.5,0) {$Z^{-1}$};
    \end{tikzpicture} &\mapsto
    \begin{tikzpicture}[baseline={([yshift=-.8ex]current bounding box.center)}]
        \draw (0,0) -- (0,1); \filldraw[black] (0,0.5) circle (2pt); \node[anchor=east] at (0,0.5) {$X$}; \node[anchor=west] at (0,0.5) {$i$};
    \end{tikzpicture}.
\end{aligned}
\end{equation}
Thus, the bulk symmetry action induces the boundary duality transformation, illustrating the correspondence between bulk symmetry and boundary duality.

\begin{figure}[htpb]
    \centering
    
    \includegraphics[]{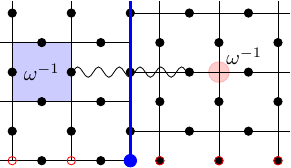}

    \vspace{10pt}
    
    \includegraphics[]{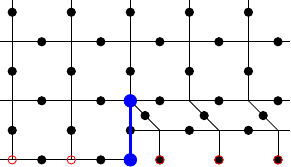}
    
    \caption{\textit{Top panel}: Correspondence between the bulk electric-magnetic duality defect (blue line) and the boundary duality defect (blue dot). When an electric excitation crosses the duality defect line, it transforms into a magnetic excitation, and vice versa. Red circles at the bottom represent boundary degrees of freedom. \textit{Bottom panel}: Endpoint of the electric-magnetic duality defect in bulk.}
    \label{fig:duality_defects}
\end{figure}

Due to the correspondence, the endpoint of the bulk symmetry defect can be interpreted as a boundary duality defect, as illustrated in Fig.~\ref{fig:duality_defects}. The red circles in the figure represent fictitious boundary degrees of freedom, which can be understood as the qudits of either the direct or dual generalized $N$-state clock model. When a bulk magnetic excitation passes through the bulk symmetry defect, it transforms into an electric excitation, and vice versa. Consequently, the bulk symmetry corresponds to the electric-magnetic self-duality. This bulk electric-magnetic duality, when constrained to the boundary Hilbert space, generates the duality in the generalized $N$-state clock model Eq.~\eqref{1d_model}.

The ordinary $\mathbb{Z}_N$ toric code is a special case of the generalized $\mathbb{Z}_N$ toric code, obtained by setting the parameters to $a_1 = a_2 = 1$. In these models, the ordinary $N$-state clock models, restricted to constrained Hilbert spaces, can be realized as boundary theories. These clock models exhibit a well-known Kramers-Wannier-like duality~\cite{whitsitt2018quantum}, which can also be understood as the boundary effect of the bulk electric-magnetic self-duality.

\section{Conclusion}\label{sec:conclusions}

In this work, we analyzed generalized $N$-state clock models, which include the transverse-field Ising model and standard $N$-state clock models as specific cases. These models display a spatially modulated symmetry if and only if $\gcd(a^L-1, N) \neq 1$ and exhibit a Kramers-Wannier-like self-duality. Interestingly, this duality becomes non-invertible precisely when the models exhibit spatially modulated symmetry, underscoring a deep connection between symmetry and duality. The non-invertible duality can be rendered invertible by restricting the Hilbert space to the charge-zero sector of the spatially modulated symmetry. However, because the restricted Hilbert space cannot be decomposed as a tensor product of local spaces, these models can no longer be realized within one spatial dimension. In other words, the restricted systems exhibit a non-invertible gravitational anomaly.

A system with a non-invertible gravitational anomaly can be realized as a lattice model by coupling to a topologically ordered system in one higher dimension. We found that the bulk conservation law in the generalized $\mathbb{Z}_N$ toric code imposes a boundary constraint identical to the condition required on the Hilbert space of the generalized $N$-state clock model to make its non-invertible duality invertible. This correspondence suggests that the bulk topological order associated with the generalized $N$-state model with spatially modulated symmetry is the generalized $\mathbb{Z}_N$ toric code. Within this framework, the boundary duality transformation corresponds to the boundary action of a bulk symmetry transformation. We also demonstrated that the relevant bulk symmetry is the electric-magnetic self-duality, with the endpoint of the bulk symmetry defect realizing the boundary duality defect.

\section*{Acknowledgements}

We thank Sungwoo Hong and Frank Verstreate for helpful discussions and thank \"Omer Mert Aksoy for bringing Refs.~\cite{pace2024gauging,pace2024spt} to our attention. D.S. and G.Y.C. are supported by Samsung Science and Technology Foundation under Project Number SSTF-BA2002-05 and SSTF-BA2401-03, the NRF of Korea (Grants No.~RS-2023-00208291, No.~2023M3K5A1094810, No.~2023M3K5A1094813, No.~RS-2024-00410027, No.~RS-2024-00444725) funded by the Korean Government (MSIT), the Air Force Office of Scientific Research under Award No.~FA2386-22-1-4061, and Institute of Basic Science under project code IBS-R014-D1.  R.-J.~S. acknowledges funding from a New Investigator Award, EPSRC grant EP/W00187X/1, a EPSRC ERC underwrite grant EP/X025829/1 as well as Trinity College, Cambridge. 

\bibliographystyle{apsrev4-2}
\bibliography{ref}

\begin{thebibliography}{38}%
\makeatletter
\providecommand \@ifxundefined [1]{%
 \@ifx{#1\undefined}
}%
\providecommand \@ifnum [1]{%
 \ifnum #1\expandafter \@firstoftwo
 \else \expandafter \@secondoftwo
 \fi
}%
\providecommand \@ifx [1]{%
 \ifx #1\expandafter \@firstoftwo
 \else \expandafter \@secondoftwo
 \fi
}%
\providecommand \natexlab [1]{#1}%
\providecommand \enquote  [1]{``#1''}%
\providecommand \bibnamefont  [1]{#1}%
\providecommand \bibfnamefont [1]{#1}%
\providecommand \citenamefont [1]{#1}%
\providecommand \href@noop [0]{\@secondoftwo}%
\providecommand \href [0]{\begingroup \@sanitize@url \@href}%
\providecommand \@href[1]{\@@startlink{#1}\@@href}%
\providecommand \@@href[1]{\endgroup#1\@@endlink}%
\providecommand \@sanitize@url [0]{\catcode `\\12\catcode `\$12\catcode `\&12\catcode `\#12\catcode `\^12\catcode `\_12\catcode `\%12\relax}%
\providecommand \@@startlink[1]{}%
\providecommand \@@endlink[0]{}%
\providecommand \url  [0]{\begingroup\@sanitize@url \@url }%
\providecommand \@url [1]{\endgroup\@href {#1}{\urlprefix }}%
\providecommand \urlprefix  [0]{URL }%
\providecommand \Eprint [0]{\href }%
\providecommand \doibase [0]{https://doi.org/}%
\providecommand \selectlanguage [0]{\@gobble}%
\providecommand \bibinfo  [0]{\@secondoftwo}%
\providecommand \bibfield  [0]{\@secondoftwo}%
\providecommand \translation [1]{[#1]}%
\providecommand \BibitemOpen [0]{}%
\providecommand \bibitemStop [0]{}%
\providecommand \bibitemNoStop [0]{.\EOS\space}%
\providecommand \EOS [0]{\spacefactor3000\relax}%
\providecommand \BibitemShut  [1]{\csname bibitem#1\endcsname}%
\let\auto@bib@innerbib\@empty
\bibitem [{\citenamefont {Kramers}\ and\ \citenamefont {Wannier}(1941)}]{kramers1941statistics}%
  \BibitemOpen
  \bibfield  {author} {\bibinfo {author} {\bibfnamefont {H.~A.}\ \bibnamefont {Kramers}}\ and\ \bibinfo {author} {\bibfnamefont {G.~H.}\ \bibnamefont {Wannier}},\ }\href {https://doi.org/10.1103/PhysRev.60.252} {\bibfield  {journal} {\bibinfo  {journal} {Physical Review}\ }\textbf {\bibinfo {volume} {60}},\ \bibinfo {pages} {252} (\bibinfo {year} {1941})}\BibitemShut {NoStop}%
\bibitem [{\citenamefont {Kogut}(1979)}]{kogut1979introduction}%
  \BibitemOpen
  \bibfield  {author} {\bibinfo {author} {\bibfnamefont {J.~B.}\ \bibnamefont {Kogut}},\ }\href {https://doi.org/10.1103/RevModPhys.51.659} {\bibfield  {journal} {\bibinfo  {journal} {Review of Modern Physics}\ }\textbf {\bibinfo {volume} {51}},\ \bibinfo {pages} {659} (\bibinfo {year} {1979})}\BibitemShut {NoStop}%
\bibitem [{\citenamefont {Witten}(1998)}]{witten1998anti}%
  \BibitemOpen
  \bibfield  {author} {\bibinfo {author} {\bibfnamefont {E.}~\bibnamefont {Witten}},\ }\href {https://doi.org/10.4310/ATMP.1998.v2.n2.a2} {\bibfield  {journal} {\bibinfo  {journal} {Advances in Theoretical and Mathematical Physics}\ }\textbf {\bibinfo {volume} {2}},\ \bibinfo {pages} {253} (\bibinfo {year} {1998})}\BibitemShut {NoStop}%
\bibitem [{\citenamefont {Maldacena}(1999)}]{maldacena1999large}%
  \BibitemOpen
  \bibfield  {author} {\bibinfo {author} {\bibfnamefont {J.}~\bibnamefont {Maldacena}},\ }\href {https://doi.org/10.1023/a:1026654312961} {\bibfield  {journal} {\bibinfo  {journal} {International Journal of Theoretical Physics}\ }\textbf {\bibinfo {volume} {38}},\ \bibinfo {pages} {1113} (\bibinfo {year} {1999})}\BibitemShut {NoStop}%
\bibitem [{\citenamefont {Seiberg}\ \emph {et~al.}(2016)\citenamefont {Seiberg}, \citenamefont {Senthil}, \citenamefont {Wang},\ and\ \citenamefont {Witten}}]{seiberg2016duality}%
  \BibitemOpen
  \bibfield  {author} {\bibinfo {author} {\bibfnamefont {N.}~\bibnamefont {Seiberg}}, \bibinfo {author} {\bibfnamefont {T.}~\bibnamefont {Senthil}}, \bibinfo {author} {\bibfnamefont {C.}~\bibnamefont {Wang}},\ and\ \bibinfo {author} {\bibfnamefont {E.}~\bibnamefont {Witten}},\ }\href {https://doi.org/10.1016/j.aop.2016.08.007} {\bibfield  {journal} {\bibinfo  {journal} {Annals of Physics}\ }\textbf {\bibinfo {volume} {374}},\ \bibinfo {pages} {395} (\bibinfo {year} {2016})}\BibitemShut {NoStop}%
\bibitem [{\citenamefont {Beekman}\ \emph {et~al.}(2017)\citenamefont {Beekman}, \citenamefont {Nissinen}, \citenamefont {Wu}, \citenamefont {Liu}, \citenamefont {Slager}, \citenamefont {Nussinov}, \citenamefont {Cvetkovic},\ and\ \citenamefont {Zaanen}}]{beekman2017dual}%
  \BibitemOpen
  \bibfield  {author} {\bibinfo {author} {\bibfnamefont {A.~J.}\ \bibnamefont {Beekman}}, \bibinfo {author} {\bibfnamefont {J.}~\bibnamefont {Nissinen}}, \bibinfo {author} {\bibfnamefont {K.}~\bibnamefont {Wu}}, \bibinfo {author} {\bibfnamefont {K.}~\bibnamefont {Liu}}, \bibinfo {author} {\bibfnamefont {R.-J.}\ \bibnamefont {Slager}}, \bibinfo {author} {\bibfnamefont {Z.}~\bibnamefont {Nussinov}}, \bibinfo {author} {\bibfnamefont {V.}~\bibnamefont {Cvetkovic}},\ and\ \bibinfo {author} {\bibfnamefont {J.}~\bibnamefont {Zaanen}},\ }\href {https://doi.org/10.1016/j.physrep.2017.03.004} {\bibfield  {journal} {\bibinfo  {journal} {Physics Report}\ }\textbf {\bibinfo {volume} {683}},\ \bibinfo {pages} {1} (\bibinfo {year} {2017})}\BibitemShut {NoStop}%
\bibitem [{\citenamefont {R\"uegg}\ \emph {et~al.}(2024)\citenamefont {R\"uegg}, \citenamefont {Chaudhary},\ and\ \citenamefont {Slager}}]{ruegg2024dualities}%
  \BibitemOpen
  \bibfield  {author} {\bibinfo {author} {\bibfnamefont {L.}~\bibnamefont {R\"uegg}}, \bibinfo {author} {\bibfnamefont {G.}~\bibnamefont {Chaudhary}},\ and\ \bibinfo {author} {\bibfnamefont {R.-J.}\ \bibnamefont {Slager}},\ }\href {https://doi.org/10.1103/PhysRevLett.133.156502} {\bibfield  {journal} {\bibinfo  {journal} {Physical Review Letters}\ }\textbf {\bibinfo {volume} {133}},\ \bibinfo {pages} {156502} (\bibinfo {year} {2024})}\BibitemShut {NoStop}%
\bibitem [{\citenamefont {Gaiotto}\ \emph {et~al.}(2015)\citenamefont {Gaiotto}, \citenamefont {Kapustin},\ and\ \citenamefont {Seiberg}}]{gaiotto2015generalized}%
  \BibitemOpen
  \bibfield  {author} {\bibinfo {author} {\bibfnamefont {D.}~\bibnamefont {Gaiotto}}, \bibinfo {author} {\bibfnamefont {A.}~\bibnamefont {Kapustin}},\ and\ \bibinfo {author} {\bibfnamefont {N.}~\bibnamefont {Seiberg}},\ }\href {https://doi.org/10.1007/JHEP02(2015)172} {\bibfield  {journal} {\bibinfo  {journal} {Journal of High Energy Physics}\ }\textbf {\bibinfo {volume} {2015}},\ \bibinfo {pages} {172} (\bibinfo {year} {2015})}\BibitemShut {NoStop}%
\bibitem [{\citenamefont {Gomes}(2023)}]{gomes2023introduction}%
  \BibitemOpen
  \bibfield  {author} {\bibinfo {author} {\bibfnamefont {P.~R.~S.}\ \bibnamefont {Gomes}},\ }\href {https://doi.org/10.21468/SciPostPhysLectNotes.74} {\bibfield  {journal} {\bibinfo  {journal} {SciPost Physics Lecture Notes}\ }\textbf {\bibinfo {volume} {74}},\ \bibinfo {pages} {1} (\bibinfo {year} {2023})}\BibitemShut {NoStop}%
\bibitem [{\citenamefont {Bhardwaj}\ \emph {et~al.}(2024)\citenamefont {Bhardwaj}, \citenamefont {Bottini}, \citenamefont {Fraser-Taliente}, \citenamefont {Gladden}, \citenamefont {Gould}, \citenamefont {Platschorre},\ and\ \citenamefont {Tillim}}]{bhardwaj2024lectures}%
  \BibitemOpen
  \bibfield  {author} {\bibinfo {author} {\bibfnamefont {L.}~\bibnamefont {Bhardwaj}}, \bibinfo {author} {\bibfnamefont {L.~E.}\ \bibnamefont {Bottini}}, \bibinfo {author} {\bibfnamefont {L.}~\bibnamefont {Fraser-Taliente}}, \bibinfo {author} {\bibfnamefont {L.}~\bibnamefont {Gladden}}, \bibinfo {author} {\bibfnamefont {D.~S.~W.}\ \bibnamefont {Gould}}, \bibinfo {author} {\bibfnamefont {A.}~\bibnamefont {Platschorre}},\ and\ \bibinfo {author} {\bibfnamefont {H.}~\bibnamefont {Tillim}},\ }\href {https://doi.org/10.1016/j.physrep.2023.11.002} {\bibfield  {journal} {\bibinfo  {journal} {Physics Report}\ }\textbf {\bibinfo {volume} {1051}},\ \bibinfo {pages} {1} (\bibinfo {year} {2024})}\BibitemShut {NoStop}%
\bibitem [{\citenamefont {Luo}\ \emph {et~al.}(2024)\citenamefont {Luo}, \citenamefont {Wang},\ and\ \citenamefont {Wang}}]{luo2024lecture}%
  \BibitemOpen
  \bibfield  {author} {\bibinfo {author} {\bibfnamefont {R.}~\bibnamefont {Luo}}, \bibinfo {author} {\bibfnamefont {Q.~R.}\ \bibnamefont {Wang}},\ and\ \bibinfo {author} {\bibfnamefont {Y.~N.}\ \bibnamefont {Wang}},\ }\href {https://doi.org/10.1016/j.physrep.2024.02.002} {\bibfield  {journal} {\bibinfo  {journal} {Physics Report}\ }\textbf {\bibinfo {volume} {1065}},\ \bibinfo {pages} {1} (\bibinfo {year} {2024})}\BibitemShut {NoStop}%
\bibitem [{\citenamefont {Shao}(2024)}]{shao2024what}%
  \BibitemOpen
  \bibfield  {author} {\bibinfo {author} {\bibfnamefont {S.-H.}\ \bibnamefont {Shao}},\ }\href {https://arxiv.org/abs/2308.00747} {\bibinfo {title} {What's done cannot be undone: Tasi lectures on non-invertible symmetries}} (\bibinfo {year} {2024}),\ \Eprint {https://arxiv.org/abs/2308.00747} {arXiv:2308.00747 [hep-th]} \BibitemShut {NoStop}%
\bibitem [{\citenamefont {Schäfer-Nameki}(2024)}]{schafernameki2024ictp}%
  \BibitemOpen
  \bibfield  {author} {\bibinfo {author} {\bibfnamefont {S.}~\bibnamefont {Schäfer-Nameki}},\ }\href {https://doi.org/10.1016/j.physrep.2024.01.007} {\bibfield  {journal} {\bibinfo  {journal} {Physics Report}\ }\textbf {\bibinfo {volume} {1063}},\ \bibinfo {pages} {1} (\bibinfo {year} {2024})}\BibitemShut {NoStop}%
\bibitem [{\citenamefont {Lootens}\ \emph {et~al.}(2023{\natexlab{a}})\citenamefont {Lootens}, \citenamefont {Delcamp}, \citenamefont {Ortiz},\ and\ \citenamefont {Verstraete}}]{lootens2023dualities}%
  \BibitemOpen
  \bibfield  {author} {\bibinfo {author} {\bibfnamefont {L.}~\bibnamefont {Lootens}}, \bibinfo {author} {\bibfnamefont {C.}~\bibnamefont {Delcamp}}, \bibinfo {author} {\bibfnamefont {G.}~\bibnamefont {Ortiz}},\ and\ \bibinfo {author} {\bibfnamefont {F.}~\bibnamefont {Verstraete}},\ }\href {https://doi.org/10.1103/PRXQuantum.4.020357} {\bibfield  {journal} {\bibinfo  {journal} {PRX Quantum}\ }\textbf {\bibinfo {volume} {4}},\ \bibinfo {pages} {020357} (\bibinfo {year} {2023}{\natexlab{a}})}\BibitemShut {NoStop}%
\bibitem [{\citenamefont {Lootens}\ \emph {et~al.}(2023{\natexlab{b}})\citenamefont {Lootens}, \citenamefont {Delcamp}, \citenamefont {Williamson},\ and\ \citenamefont {Verstraete}}]{lootens2023low}%
  \BibitemOpen
  \bibfield  {author} {\bibinfo {author} {\bibfnamefont {L.}~\bibnamefont {Lootens}}, \bibinfo {author} {\bibfnamefont {C.}~\bibnamefont {Delcamp}}, \bibinfo {author} {\bibfnamefont {D.}~\bibnamefont {Williamson}},\ and\ \bibinfo {author} {\bibfnamefont {F.}~\bibnamefont {Verstraete}},\ }\href {https://arxiv.org/abs/2311.01439} {\bibinfo {title} {Low-depth unitary quantum circuits for dualities in one-dimensional quantum lattice models}} (\bibinfo {year} {2023}{\natexlab{b}}),\ \Eprint {https://arxiv.org/abs/2311.01439} {arXiv:2311.01439 [quant-ph]} \BibitemShut {NoStop}%
\bibitem [{\citenamefont {Lootens}\ \emph {et~al.}(2024)\citenamefont {Lootens}, \citenamefont {Delcamp},\ and\ \citenamefont {Verstraete}}]{lootens2024dualities}%
  \BibitemOpen
  \bibfield  {author} {\bibinfo {author} {\bibfnamefont {L.}~\bibnamefont {Lootens}}, \bibinfo {author} {\bibfnamefont {C.}~\bibnamefont {Delcamp}},\ and\ \bibinfo {author} {\bibfnamefont {F.}~\bibnamefont {Verstraete}},\ }\href {https://doi.org/10.1103/PRXQuantum.5.010338} {\bibfield  {journal} {\bibinfo  {journal} {PRX Quantum}\ }\textbf {\bibinfo {volume} {5}},\ \bibinfo {pages} {010338} (\bibinfo {year} {2024})}\BibitemShut {NoStop}%
\bibitem [{\citenamefont {Moradi}\ \emph {et~al.}(2023)\citenamefont {Moradi}, \citenamefont {Moosavian},\ and\ \citenamefont {Tiwari}}]{moradi2023topological}%
  \BibitemOpen
  \bibfield  {author} {\bibinfo {author} {\bibfnamefont {H.}~\bibnamefont {Moradi}}, \bibinfo {author} {\bibfnamefont {S.~F.}\ \bibnamefont {Moosavian}},\ and\ \bibinfo {author} {\bibfnamefont {A.}~\bibnamefont {Tiwari}},\ }\href {https://doi.org/10.21468/SciPostPhysCore.6.4.066} {\bibfield  {journal} {\bibinfo  {journal} {SciPost Physics Core}\ }\textbf {\bibinfo {volume} {6}},\ \bibinfo {pages} {066} (\bibinfo {year} {2023})}\BibitemShut {NoStop}%
\bibitem [{\citenamefont {Chatterjee}\ and\ \citenamefont {Wen}(2023{\natexlab{a}})}]{chatterjee2023symmetry}%
  \BibitemOpen
  \bibfield  {author} {\bibinfo {author} {\bibfnamefont {A.}~\bibnamefont {Chatterjee}}\ and\ \bibinfo {author} {\bibfnamefont {X.-G.}\ \bibnamefont {Wen}},\ }\href {https://doi.org/10.1103/PhysRevB.107.155136} {\bibfield  {journal} {\bibinfo  {journal} {Physical Review B}\ }\textbf {\bibinfo {volume} {107}},\ \bibinfo {pages} {155136} (\bibinfo {year} {2023}{\natexlab{a}})}\BibitemShut {NoStop}%
\bibitem [{\citenamefont {Chatterjee}\ and\ \citenamefont {Wen}(2023{\natexlab{b}})}]{chatterjee2023holographic}%
  \BibitemOpen
  \bibfield  {author} {\bibinfo {author} {\bibfnamefont {A.}~\bibnamefont {Chatterjee}}\ and\ \bibinfo {author} {\bibfnamefont {X.-G.}\ \bibnamefont {Wen}},\ }\href {https://doi.org/10.1103/PhysRevB.108.075105} {\bibfield  {journal} {\bibinfo  {journal} {Physical Review B}\ }\textbf {\bibinfo {volume} {108}},\ \bibinfo {pages} {075105} (\bibinfo {year} {2023}{\natexlab{b}})}\BibitemShut {NoStop}%
\bibitem [{\citenamefont {Inamura}\ and\ \citenamefont {Wen}(2023)}]{inamura2023symmetry}%
  \BibitemOpen
  \bibfield  {author} {\bibinfo {author} {\bibfnamefont {K.}~\bibnamefont {Inamura}}\ and\ \bibinfo {author} {\bibfnamefont {X.-G.}\ \bibnamefont {Wen}},\ }\href {https://arxiv.org/abs/2310.05790} {\bibinfo {title} {2+1d symmetry-topological-order from local symmetric operators in 1+1d}} (\bibinfo {year} {2023}),\ \Eprint {https://arxiv.org/abs/2310.05790} {arXiv:2310.05790 [cond-mat.str-el]} \BibitemShut {NoStop}%
\bibitem [{\citenamefont {Vanhove}\ \emph {et~al.}(2024)\citenamefont {Vanhove}, \citenamefont {Ravindran}, \citenamefont {Stephen}, \citenamefont {Wen},\ and\ \citenamefont {Chen}}]{vanhove2024duality}%
  \BibitemOpen
  \bibfield  {author} {\bibinfo {author} {\bibfnamefont {R.}~\bibnamefont {Vanhove}}, \bibinfo {author} {\bibfnamefont {V.}~\bibnamefont {Ravindran}}, \bibinfo {author} {\bibfnamefont {D.~T.}\ \bibnamefont {Stephen}}, \bibinfo {author} {\bibfnamefont {X.-G.}\ \bibnamefont {Wen}},\ and\ \bibinfo {author} {\bibfnamefont {X.}~\bibnamefont {Chen}},\ }\href {https://arxiv.org/abs/2409.06647} {\bibinfo {title} {Duality via sequential quantum circuit in the topological holography formalism}} (\bibinfo {year} {2024}),\ \Eprint {https://arxiv.org/abs/2409.06647} {arXiv:2409.06647 [cond-mat.str-el]} \BibitemShut {NoStop}%
\bibitem [{\citenamefont {Sala}\ \emph {et~al.}(2022)\citenamefont {Sala}, \citenamefont {Lehmann}, \citenamefont {Rakovszky},\ and\ \citenamefont {Pollmann}}]{sala2022dynamics}%
  \BibitemOpen
  \bibfield  {author} {\bibinfo {author} {\bibfnamefont {P.}~\bibnamefont {Sala}}, \bibinfo {author} {\bibfnamefont {J.}~\bibnamefont {Lehmann}}, \bibinfo {author} {\bibfnamefont {T.}~\bibnamefont {Rakovszky}},\ and\ \bibinfo {author} {\bibfnamefont {F.}~\bibnamefont {Pollmann}},\ }\href {https://doi.org/10.1103/PhysRevLett.129.170601} {\bibfield  {journal} {\bibinfo  {journal} {Physical Review Letters}\ }\textbf {\bibinfo {volume} {129}},\ \bibinfo {pages} {170601} (\bibinfo {year} {2022})}\BibitemShut {NoStop}%
\bibitem [{\citenamefont {Delfino}\ \emph {et~al.}(2023)\citenamefont {Delfino}, \citenamefont {Chamon},\ and\ \citenamefont {You}}]{delfino2023fractons}%
  \BibitemOpen
  \bibfield  {author} {\bibinfo {author} {\bibfnamefont {G.}~\bibnamefont {Delfino}}, \bibinfo {author} {\bibfnamefont {C.}~\bibnamefont {Chamon}},\ and\ \bibinfo {author} {\bibfnamefont {Y.}~\bibnamefont {You}},\ }\href {https://arxiv.org/abs/2306.17121} {\bibinfo {title} {{2D Fractons from Gauging Exponential Symmetries}}} (\bibinfo {year} {2023}),\ \Eprint {https://arxiv.org/abs/2306.17121} {arXiv:2306.17121 [cond-mat.str-el]} \BibitemShut {NoStop}%
\bibitem [{\citenamefont {Han}\ \emph {et~al.}(2024)\citenamefont {Han}, \citenamefont {Lake}, \citenamefont {Lam}, \citenamefont {Verresen},\ and\ \citenamefont {You}}]{han2024topological}%
  \BibitemOpen
  \bibfield  {author} {\bibinfo {author} {\bibfnamefont {J.~H.}\ \bibnamefont {Han}}, \bibinfo {author} {\bibfnamefont {E.}~\bibnamefont {Lake}}, \bibinfo {author} {\bibfnamefont {H.~T.}\ \bibnamefont {Lam}}, \bibinfo {author} {\bibfnamefont {R.}~\bibnamefont {Verresen}},\ and\ \bibinfo {author} {\bibfnamefont {Y.}~\bibnamefont {You}},\ }\href {https://doi.org/10.1103/PhysRevB.109.125121} {\bibfield  {journal} {\bibinfo  {journal} {Physical Review B}\ }\textbf {\bibinfo {volume} {109}},\ \bibinfo {pages} {125121} (\bibinfo {year} {2024})}\BibitemShut {NoStop}%
\bibitem [{\citenamefont {Lam}(2024)}]{lam2024classification}%
  \BibitemOpen
  \bibfield  {author} {\bibinfo {author} {\bibfnamefont {H.~T.}\ \bibnamefont {Lam}},\ }\href {https://doi.org/10.1103/PhysRevB.109.115142} {\bibfield  {journal} {\bibinfo  {journal} {Physical Review B}\ }\textbf {\bibinfo {volume} {109}},\ \bibinfo {pages} {115142} (\bibinfo {year} {2024})}\BibitemShut {NoStop}%
\bibitem [{\citenamefont {Sala}\ \emph {et~al.}(2024)\citenamefont {Sala}, \citenamefont {You}, \citenamefont {Hauschild},\ and\ \citenamefont {Motrunich}}]{sala2024exotic}%
  \BibitemOpen
  \bibfield  {author} {\bibinfo {author} {\bibfnamefont {P.}~\bibnamefont {Sala}}, \bibinfo {author} {\bibfnamefont {Y.}~\bibnamefont {You}}, \bibinfo {author} {\bibfnamefont {J.}~\bibnamefont {Hauschild}},\ and\ \bibinfo {author} {\bibfnamefont {O.}~\bibnamefont {Motrunich}},\ }\href {https://doi.org/10.1103/PhysRevB.109.014406} {\bibfield  {journal} {\bibinfo  {journal} {Physical Review B}\ }\textbf {\bibinfo {volume} {109}},\ \bibinfo {pages} {014406} (\bibinfo {year} {2024})}\BibitemShut {NoStop}%
\bibitem [{\citenamefont {Pace}\ \emph {et~al.}(2024{\natexlab{a}})\citenamefont {Pace}, \citenamefont {Delfino}, \citenamefont {Lam},\ and\ \citenamefont {Ömer M.~Aksoy}}]{pace2024gauging}%
  \BibitemOpen
  \bibfield  {author} {\bibinfo {author} {\bibfnamefont {S.~D.}\ \bibnamefont {Pace}}, \bibinfo {author} {\bibfnamefont {G.}~\bibnamefont {Delfino}}, \bibinfo {author} {\bibfnamefont {H.~T.}\ \bibnamefont {Lam}},\ and\ \bibinfo {author} {\bibnamefont {Ömer M.~Aksoy}},\ }\href {https://arxiv.org/abs/2406.12962} {\bibinfo {title} {{Gauging modulated symmetries: Kramers-Wannier dualities and non-invertible reflections}}} (\bibinfo {year} {2024}{\natexlab{a}}),\ \Eprint {https://arxiv.org/abs/2406.12962} {arXiv:2406.12962 [cond-mat.str-el]} \BibitemShut {NoStop}%
\bibitem [{\citenamefont {Pace}\ \emph {et~al.}(2024{\natexlab{b}})\citenamefont {Pace}, \citenamefont {Lam},\ and\ \citenamefont {Ömer M.~Aksoy}}]{pace2024spt}%
  \BibitemOpen
  \bibfield  {author} {\bibinfo {author} {\bibfnamefont {S.~D.}\ \bibnamefont {Pace}}, \bibinfo {author} {\bibfnamefont {H.~T.}\ \bibnamefont {Lam}},\ and\ \bibinfo {author} {\bibnamefont {Ömer M.~Aksoy}},\ }\href {https://arxiv.org/abs/2409.18113} {\bibinfo {title} {{(SPT-)LSM theorems from projective non-invertible symmetries}}} (\bibinfo {year} {2024}{\natexlab{b}}),\ \Eprint {https://arxiv.org/abs/2409.18113} {arXiv:2409.18113 [cond-mat.str-el]} \BibitemShut {NoStop}%
\bibitem [{\citenamefont {Hu}\ and\ \citenamefont {Watanabe}(2023)}]{hu2023spontaneous}%
  \BibitemOpen
  \bibfield  {author} {\bibinfo {author} {\bibfnamefont {Y.}~\bibnamefont {Hu}}\ and\ \bibinfo {author} {\bibfnamefont {H.}~\bibnamefont {Watanabe}},\ }\href {https://doi.org/10.1103/PhysRevB.107.195139} {\bibfield  {journal} {\bibinfo  {journal} {Physical Review B}\ }\textbf {\bibinfo {volume} {107}},\ \bibinfo {pages} {195139} (\bibinfo {year} {2023})}\BibitemShut {NoStop}%
\bibitem [{\citenamefont {Ji}\ and\ \citenamefont {Wen}(2022)}]{ji2022unified}%
  \BibitemOpen
  \bibfield  {author} {\bibinfo {author} {\bibfnamefont {W.}~\bibnamefont {Ji}}\ and\ \bibinfo {author} {\bibfnamefont {X.-G.}\ \bibnamefont {Wen}},\ }\href {https://arxiv.org/abs/2106.02069} {\bibinfo {title} {A unified view on symmetry, anomalous symmetry and non-invertible gravitational anomaly}} (\bibinfo {year} {2022}),\ \Eprint {https://arxiv.org/abs/2106.02069} {arXiv:2106.02069 [cond-mat.str-el]} \BibitemShut {NoStop}%
\bibitem [{\citenamefont {Watanabe}\ \emph {et~al.}(2023)\citenamefont {Watanabe}, \citenamefont {Cheng},\ and\ \citenamefont {Fuji}}]{watanabe2023ground}%
  \BibitemOpen
  \bibfield  {author} {\bibinfo {author} {\bibfnamefont {H.}~\bibnamefont {Watanabe}}, \bibinfo {author} {\bibfnamefont {M.}~\bibnamefont {Cheng}},\ and\ \bibinfo {author} {\bibfnamefont {Y.}~\bibnamefont {Fuji}},\ }\href {https://doi.org/10.1063/5.0134010} {\bibfield  {journal} {\bibinfo  {journal} {Journal of Mathematical Physics}\ }\textbf {\bibinfo {volume} {64}},\ \bibinfo {pages} {051901} (\bibinfo {year} {2023})}\BibitemShut {NoStop}%
\bibitem [{\citenamefont {Minwalla}\ \emph {et~al.}(2000)\citenamefont {Minwalla}, \citenamefont {Raamsdonk},\ and\ \citenamefont {Seiberg}}]{minwalla2000noncommutative}%
  \BibitemOpen
  \bibfield  {author} {\bibinfo {author} {\bibfnamefont {S.}~\bibnamefont {Minwalla}}, \bibinfo {author} {\bibfnamefont {M.~V.}\ \bibnamefont {Raamsdonk}},\ and\ \bibinfo {author} {\bibfnamefont {N.}~\bibnamefont {Seiberg}},\ }\href {https://doi.org/10.1088/1126-6708/2000/02/020} {\bibfield  {journal} {\bibinfo  {journal} {Journal of High Energy Physics}\ }\textbf {\bibinfo {volume} {2000}},\ \bibinfo {pages} {2} (\bibinfo {year} {2000})}\BibitemShut {NoStop}%
\bibitem [{\citenamefont {Gorantla}\ \emph {et~al.}(2021)\citenamefont {Gorantla}, \citenamefont {Lam}, \citenamefont {Seiberg},\ and\ \citenamefont {Shao}}]{gorantla2021low}%
  \BibitemOpen
  \bibfield  {author} {\bibinfo {author} {\bibfnamefont {P.}~\bibnamefont {Gorantla}}, \bibinfo {author} {\bibfnamefont {H.~T.}\ \bibnamefont {Lam}}, \bibinfo {author} {\bibfnamefont {N.}~\bibnamefont {Seiberg}},\ and\ \bibinfo {author} {\bibfnamefont {S.-H.}\ \bibnamefont {Shao}},\ }\href {https://doi.org/10.1103/PhysRevB.104.235116} {\bibfield  {journal} {\bibinfo  {journal} {Physical Review B}\ }\textbf {\bibinfo {volume} {104}},\ \bibinfo {pages} {235116} (\bibinfo {year} {2021})}\BibitemShut {NoStop}%
\bibitem [{\citenamefont {Rudelius}\ \emph {et~al.}(2021)\citenamefont {Rudelius}, \citenamefont {Seiberg},\ and\ \citenamefont {Shao}}]{rudelius2021fractons}%
  \BibitemOpen
  \bibfield  {author} {\bibinfo {author} {\bibfnamefont {T.}~\bibnamefont {Rudelius}}, \bibinfo {author} {\bibfnamefont {N.}~\bibnamefont {Seiberg}},\ and\ \bibinfo {author} {\bibfnamefont {S.-H.}\ \bibnamefont {Shao}},\ }\href {https://doi.org/10.1103/PhysRevB.103.195113} {\bibfield  {journal} {\bibinfo  {journal} {Physical Review B}\ }\textbf {\bibinfo {volume} {103}},\ \bibinfo {pages} {195113} (\bibinfo {year} {2021})}\BibitemShut {NoStop}%
\bibitem [{\citenamefont {You}\ and\ \citenamefont {Moessner}(2022)}]{you2022fractonic}%
  \BibitemOpen
  \bibfield  {author} {\bibinfo {author} {\bibfnamefont {Y.}~\bibnamefont {You}}\ and\ \bibinfo {author} {\bibfnamefont {R.}~\bibnamefont {Moessner}},\ }\href {https://doi.org/10.1103/PhysRevB.106.115145} {\bibfield  {journal} {\bibinfo  {journal} {Physical Review B}\ }\textbf {\bibinfo {volume} {106}},\ \bibinfo {pages} {115145} (\bibinfo {year} {2022})}\BibitemShut {NoStop}%
\bibitem [{\citenamefont {Kim}\ \emph {et~al.}(2024)\citenamefont {Kim}, \citenamefont {Oh}, \citenamefont {Bulmash},\ and\ \citenamefont {Han}}]{kim2024unveiling}%
  \BibitemOpen
  \bibfield  {author} {\bibinfo {author} {\bibfnamefont {J.}~\bibnamefont {Kim}}, \bibinfo {author} {\bibfnamefont {Y.-T.}\ \bibnamefont {Oh}}, \bibinfo {author} {\bibfnamefont {D.}~\bibnamefont {Bulmash}},\ and\ \bibinfo {author} {\bibfnamefont {J.~H.}\ \bibnamefont {Han}},\ }\href {https://arxiv.org/abs/2310.09425} {\bibinfo {title} {{Unveiling UV/IR Mixing via Symmetry Defects: A View from Topological Entanglement Entropy}}} (\bibinfo {year} {2024}),\ \Eprint {https://arxiv.org/abs/2310.09425} {arXiv:2310.09425 [cond-mat.str-el]} \BibitemShut {NoStop}%
\bibitem [{\citenamefont {Yan}\ and\ \citenamefont {Li}(2024)}]{yan2024generalized}%
  \BibitemOpen
  \bibfield  {author} {\bibinfo {author} {\bibfnamefont {H.}~\bibnamefont {Yan}}\ and\ \bibinfo {author} {\bibfnamefont {L.}~\bibnamefont {Li}},\ }\href {https://arxiv.org/abs/2403.16017} {\bibinfo {title} {{Generalized Kramers-Wanier Duality from Bilinear Phase Map}}} (\bibinfo {year} {2024}),\ \Eprint {https://arxiv.org/abs/2403.16017} {arXiv:2403.16017 [cond-mat.str-el]} \BibitemShut {NoStop}%
\bibitem [{\citenamefont {Whitsitt}\ \emph {et~al.}(2018)\citenamefont {Whitsitt}, \citenamefont {Samajdar},\ and\ \citenamefont {Sachdev}}]{whitsitt2018quantum}%
  \BibitemOpen
  \bibfield  {author} {\bibinfo {author} {\bibfnamefont {S.}~\bibnamefont {Whitsitt}}, \bibinfo {author} {\bibfnamefont {R.}~\bibnamefont {Samajdar}},\ and\ \bibinfo {author} {\bibfnamefont {S.}~\bibnamefont {Sachdev}},\ }\href {https://doi.org/10.1103/PhysRevB.98.205118} {\bibfield  {journal} {\bibinfo  {journal} {Physical Review B}\ }\textbf {\bibinfo {volume} {98}},\ \bibinfo {pages} {205118} (\bibinfo {year} {2018})}\BibitemShut {NoStop}%
\end{thebibliography}%
\end{document}